\documentclass{PoS}

\title{Black hole's life at colliders}

\ShortTitle{Black hole's Life at colliders}

\author{\speaker{Seong Chan Park}\thanks{We thanks to Daisuke Ida and Kin-ya Oda for valuable collaborations.}\\
        FRDP, School of astronomy and physics, Seoul National University, Seoul 151-742, Korea\\
        E-mail: \email{spark@phya.snu.ac.kr}}

\abstract{In the series of papers by Ida, Oda and Park, the complete description of Hawking radiation to the brane localized Standard Model fields from mini black holes in the low energy gravity scenarios are obtained. Here we briefly review what we have learned in those papers.}

\FullConference{From Strings to LHC\\
		January 2-10 2007\\
		International Centre Dona Paula, Goa India}

\begin{document}

\section{Introduction}
We briefly review recent developments in the mini black hole production and evaporation mainly based on the series of works done by Ida, Oda and Park \cite{Ida:2002ez,Ida:2005zi,Ida:2005ax,Ida:2006tf}. In the low energy gravity scenarios such as ADD and RS-I, the CERN Large Hadronic collider (LHC) will become a black hole factory \cite{Giddings:2001bu, Dimopoulos:2001hw}. Above the TeV Planck scale, the classical production cross section of the $(4+n)$-dimensional black hole grows geometrically $\sigma \sim \hat{s}^{1/(n+1)}$, with $\sqrt{\hat{s}}$ being the center of mass energy of the parton scattering.

Once produced, black hole loses its energy or mass primarily via Hawking (thermal) radiation. The Hawking radiation goes mainly into the standard model quarks and leptons (spinors) and gauge bosons (vectors) localized on the brane, except for a few gravitons and Higgs boson(s). The quanta of Hawking radiation will have characteristic energy spectrum determined by the Hawking temperature and the greybody factor. The process of Hawking radiation in four dimensional rotating black hole has been treated in detail by Teukolsky, Press, Page and others in 1970s'. In higher dimensions, however, it is shown that the process has quite different features.

\begin{itemize}
  \item Hawking temperaure $T_H \propto (M_{bh})/M_*)^{1/(n+1)}$ of a $(4+n)$ dimensional black hole is much higher than 4 dimensional one with the {\it small} fundamental scale $M_*\sim$ TeV $\ll M_{\rm Planck}$. With this high temperature, the number of available degrees of freedom for Hawking radiation are much bigger in $(4+n)$ dimensions with all the standard model particles localized on the brane \cite{Emparan:2000rs}.
  \item The near horizon geometry of $(4+n)$ dimensional black hole is quite complicated. Its geometry is different from that of a four dimensional Kerr black hole. With the highly modified geometry in the vicinity of the event horizon, frequency dependent correction factor of Hawking radiation, i.e., greybody factor, is also largely modified (also see the references \cite{Harris:2005jx,Casals:2005sa,Casals:2006xp} as independent studies on the same topic).
\end{itemize}

To understand the physics of those black holes, we have to understand the greybody factor of higher dimensional, rotating black hole \cite{Park:2001xc,Park:2004fj}.

\section{Generalized Teukolsky equation and greybody factor}
The induced metric on the three-brane in the $(4+n)$-dimensional
Myers-Perry solution \cite{Myers:1986un} with a single nonzero angular momentum is given by
\begin{eqnarray}
g&=&{\Delta-a^2\sin^2\vartheta\over\Sigma}dt^2
+{2a(r^2+a^2-\Delta)\sin^2\vartheta\over\Sigma}dtd\varphi\nonumber\\
&&{}-{(r^2+a^2)^2-\Delta a^2\sin^2\vartheta\over\Sigma}\sin^2\vartheta d\varphi^2
-{\Sigma\over\Delta}dr^2-\Sigma d\vartheta^2,
\label{myers-perry}
\end{eqnarray}
where
\begin{eqnarray}
\Sigma=r^2+a^2\cos^2\vartheta,~~~
\Delta=r^2+a^2-\mu r^{1-n}.
\end{eqnarray}
The parameters $\mu$ and $a$ are equivalent to the total mass $M$ and
the angular momentum $J$
\begin{equation}
M={(2+n)A_{2+n}\mu\over 16\pi G_{4+n}},~~~J={A_{2+n}\mu a\over 8\pi G_{4+n}}
\end{equation}
evaluated at the spatial infinity of the $(4+n)$-dimensional space-time,
respectively,  where
$A_{2+n}=2\pi^{(3+n)/2}/\Gamma((3+n)/2)$ is the area of a unit $(2+n)$-sphere
and $G_{4+n}$ is the $(4+n)$-dimensional Newton constant of gravitation.

1.Subtracting outgoing wave contamination at NH and separating
ingoing and outgoing wave at FF are described. 2.Here we answer the
question :what fraction of energy would be radiated into Hawking
radiation in spin-down phase.

\subsection{Asymptotic solutions in Kerr-Newman frame}
We are given a linear, second-order equation, say
\begin{eqnarray}
\frac{d^2 R}{d r^2}+{\cal \eta} \frac{d R}{d r} +{\cal \tau} R =0,
\label{diffeqKN}
\end{eqnarray}
where $\eta$ and $\tau$ are determined in Kerr-Newman frame as:
\begin{eqnarray}
\eta&=&-\frac{(s-1)\Delta'+2iK}{\Delta}, \\
\tau&=&\frac{2i \omega r(2s-1)-\lambda}{\Delta}.
\end{eqnarray}
\begin{eqnarray}
\Delta&=&r^2 + a^2
-(r_H^2+a^2)\left(\frac{r_H}{r}\right)^{n-1},\nonumber \\
K&=&(r^2+a^2)w - m a.
\end{eqnarray}

The asymptotic solutions are given at NH and FF limits:
\begin{eqnarray}
R^{\rm NH}&\sim& W_{\rm in}+W_{\rm out}e^{2 i k r_*}\Delta^s, \\
R^{\rm FF}&\sim& Y_{\rm in}r^{2s-1}+Y_{\rm out}e^{2 i k r_*}/r.
\end{eqnarray}

\subsection{BC: Subtracting outgoing contamination at NH} The
solutions near the horizon $r\rightarrow r_H$ are
\begin{eqnarray}
R^{\rm NH}_{\rm in} &=& 1+ a_1 (r-r_H) + \frac{a_2}{2}(r-r_H)^2+ \cdots, \nonumber \\
R^{\rm NH}_{\rm out} &=& e^{2 i k r_*}(r-r_H)^s\left(1+ b_1(r-r_H)
+\cdots \right),
\end{eqnarray}
where the coefficients $a_i$'s and $b_i$'s are straightforward to
compute:
\begin{eqnarray}
a_1&=& -\frac{\tau_{-1}}{\eta_{-1}}, \\
a_2&=& -\frac{(\eta_0+\tau_{-1})a_1+\tau_0}{1+\eta_{-1}},\\
&&\cdots,
\end{eqnarray}
where $\eta_j$ and $\tau_j$ are $j$-th order coefficients of Taylor
expansion  of $\eta$ and $\tau$, respectively.

For $s=1/2$ and $1$,
\begin{eqnarray}
\tau_{-1}&=&\frac{2i\omega \delta_{s,1}-\lambda}{\Delta_1},\\
\tau_0&=&\lambda\frac{\Delta_2}{\Delta_1^2}+\delta_{s,1}\frac{2i\omega}{\Delta_1^2}(\Delta_1-\Delta_2),\\
\eta_{-1}&=&\frac{1}{2}\delta_{s,1/2}-2i \frac{K_0}{\Delta_1},\\
\eta_0&=&-\frac{2i}{\Delta_1^2}(K_1\Delta_1-K_0\Delta_2),
\end{eqnarray}
where
\begin{eqnarray}
\Delta_1&=&2+(n-1)(1+a^2),\\
\Delta_2&=&1-n(n-1)(1+a^2)/2,\\
K_0&=&(1+a^2)\omega-a m,\\
K_1&=&2\omega.
\end{eqnarray}

The problem is to integrate Eq.\ref{diffeqKN} from purely ingoing
initial conditions at $r=r_0$ out to $r\rightarrow \infty$. Choosing
the positive choice for $s$ makes $Y_{\rm in}$ stable and easily
determined by an outward integration. However, for such an
integration $R^{\rm NH}_{\rm out}$ is unstable against contaminating
the purely ingoing solution. We can avoid the difficulty in
mid-integration, relying on a mathematical transformation of the
equation to stabilize the solutions in the two asymptotic regimes
$r\rightarrow r_H$ and $r\rightarrow \infty$ as follows. To
counteract the above contamination, let
\begin{eqnarray}
\tilde{R}= R-(1+ a_1 (r-r_H)).
\end{eqnarray}
Then $f$ satisfies the equation
\begin{eqnarray}
{\cal L} \tilde{R} = g, \label{diffeq2}
\end{eqnarray}
where ${\cal L}=d^2/dr^2 +\eta d/dr +\tau$ and $g=-{\cal L}\left(1+
a_1 (r-r_H)\right)=-\eta a_1-\tau\left(1+ a_1 (r-r_H)\right)$.
Equation \ref{diffeq2} is now stably integrated through both
asymptotic regimes, i.e., from the near horizon to far field
regimes.

Near the horizon, $\tilde{R}$ becomes
\begin{eqnarray}
\tilde{R}(r\rightarrow r_H) &=&\frac{a_2}{2} (r-r_H)^2+\cdots.
\end{eqnarray}

\subsection{Separating solution at FF} For $s=1/2$,

\begin{eqnarray}
R^{\rm FF}_{\rm in} &\sim& 1+ \frac{c^f_1}{r} + \frac{c^f_2}{r^2}+\frac{c^f_3}{r^3} +\cdots , \nonumber \\
R^{\rm FF}_{\rm out} &\sim& e^{2 i k r_*}\frac{1}{r} ( 1+
\frac{d^f_1}{r}+\frac{d^f_2}{r^2}\cdots),
\end{eqnarray}
where
\begin{eqnarray}
c^f_1&=&-i \frac{\lambda}{2\omega},\\
\cdots
\end{eqnarray}

Then, $\tilde{R}_{1/2}$ becomes
\begin{eqnarray}
\tilde{R}_{1/2}(r\rightarrow \infty) \simeq \left(Y_{\rm in}+1-a_1
\right) + a_1 r+\frac{Y_{\rm in} c^f_1}{r}+Y_{\rm
out}\frac{e^{2i\omega r_*}}{r}.
\end{eqnarray}

Using this expression, we can easily read out $Y$'s without
numerical difficulties.

Finally, the greybody factor could be written as
\begin{eqnarray}
\Gamma_{s=1/2} = 1-\frac{2\omega}{|c^f_1|}\frac{|Y_{\rm
out}|^2}{|Y_{\rm in}|^2}.
\end{eqnarray}

For $s=1$,
\begin{eqnarray}
R^{\rm FF}_{\rm in} &\sim& r (1+ \frac{c^v_1}{r} + \frac{c^v_2}{r^2}+\frac{c^v_3}{r^3} +\cdots) , \nonumber \\
R^{\rm FF}_{\rm out} &\sim& e^{2 i k r_*}\frac{1}{r} ( 1+
\frac{d^v_1}{r}+\frac{d^v_2}{r^2}+\frac{d^v_3}{r^3} +\cdots),
\end{eqnarray}
where
\begin{eqnarray}
c^v_1&=&-i \frac{\lambda}{2\omega},\\
c^v_2&=&-\frac{\lambda^2-4 a\omega (a\omega-m)}{8\omega^2},\\
&\cdots&.\nonumber
\end{eqnarray}

Then, $\tilde{R}_1$ becomes
\begin{eqnarray}
\tilde{R}_1(r\rightarrow \infty) \simeq \left(Y_{\rm in}-a_1
\right)r + \left(Y_{\rm in}c^v_1-1+a_1\right)+\frac{Y_{\rm in}
c^v_2}{r}+Y_{\rm out}\frac{e^{2i\omega r_*}}{r}.
\end{eqnarray}

Using this expression, we can easily read out $Y$'s without
numerical difficulties.

Finally, the greybody factor could be written as
\begin{eqnarray}
\Gamma_{s=1} = 1-\frac{2\omega^2}{|c^v_2|}\frac{|Y_{\rm
out}|^2}{|Y_{\rm in}|^2}.
\end{eqnarray}

\section{Hawking radiations of mini-black hole}
Schematically black hole evolution follows five successive steps as is depicted in Fig.1: the production phase, the balding phase, the spin-down phase, the Schwarzschild phase and the Planck phase. When a black hole is produced in high energy collision (production phase), the geometry is highly irregular, and could even be topologically non-trivial. By emitting (bulk) gravitons and other particles, the black hole will be settled down to a rotating black hole which is supposed to be well described by Myers-Perry solution in $(4+n)$ dimensional spacetime (balding phase).

The decay in spin-down and Schwarzschild phases are calculable in terms of Hawking radiation.
We are interested in those phases (spin-down
phase and Schwarzschild phase) and would  answer what
fraction of energy will be lost in each of these phases.

The rate of energy (and angular momentum) loss by Hawking radiation is given as follows:
\begin{eqnarray}
-\frac{d}{dt}\left(%
\begin{array}{c}
  M \\
  J \\
\end{array}%
\right) =\frac{1}{2\pi}\sum_{s,l,m}g_s \int_0^\infty  d \omega \langle N_{s,l,m} \rangle\left(%
\begin{array}{c}
  \omega \\
  m \\
\end{array}%
\right), \label{rates}
\end{eqnarray}
where $g_s$ is the number of ``massless" degrees of freedom at
temperature $T$, namely, the number of degrees of freedom whose
masses are smaller than $T$, with spin $s$. The expected number of
particles of the species of spin $s$ emitted in the mode with
spheroidal harmonics $l$, axial angular momentum $m$ is
\begin{eqnarray}
\langle N_{s,l,m} \rangle = \frac{{}_s\Gamma_{l,m}(a,
\omega)}{e^{(\omega-m\Omega)/T}-(-1)^{2s}}.
\end{eqnarray}
%

\begin{figure}[h]
\begin{center}
\includegraphics[width=.43\linewidth]{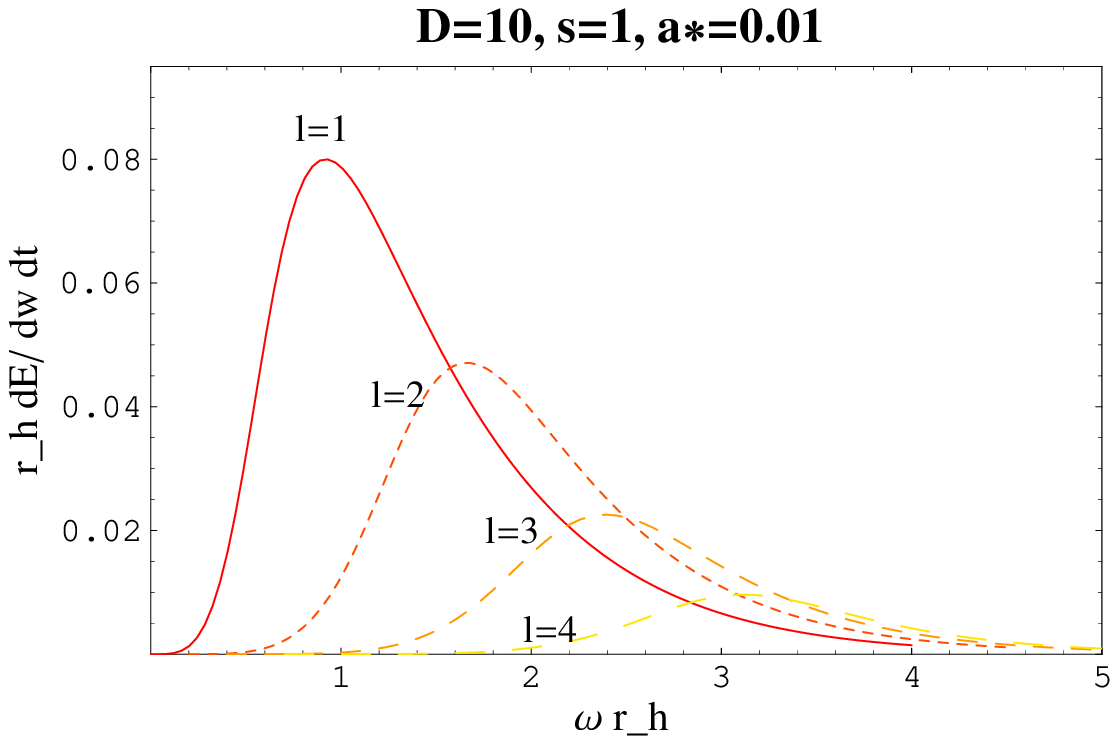}
\includegraphics[width=.43\linewidth]{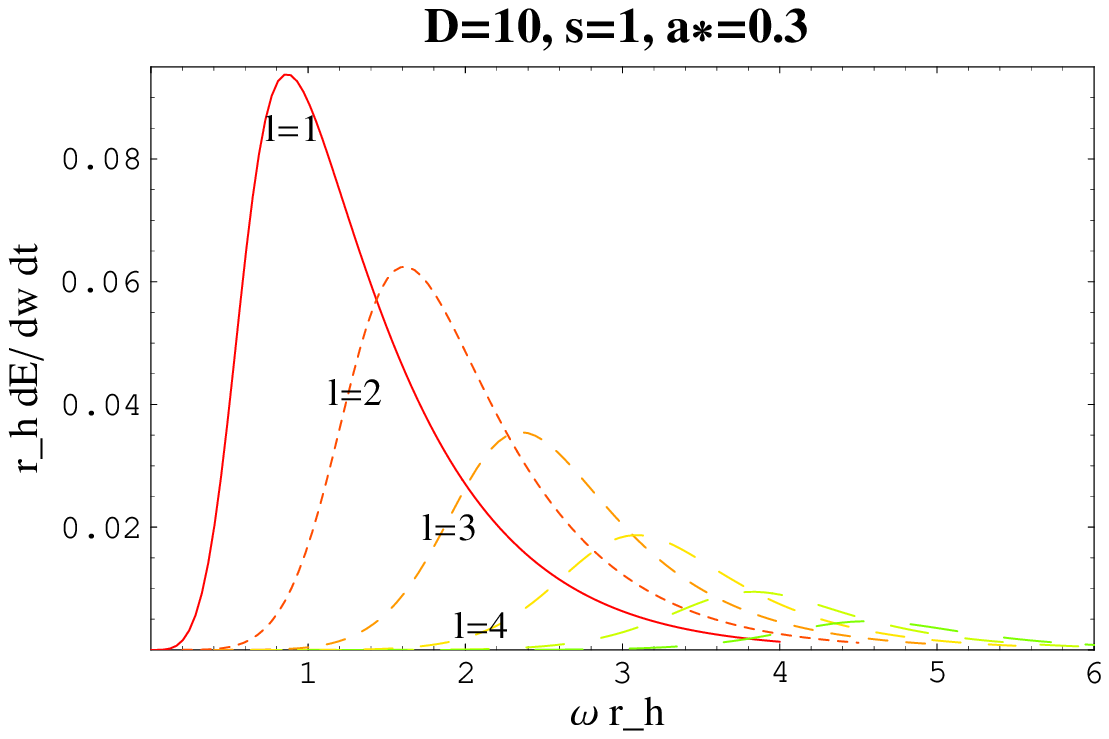}
\includegraphics[width=.43\linewidth]{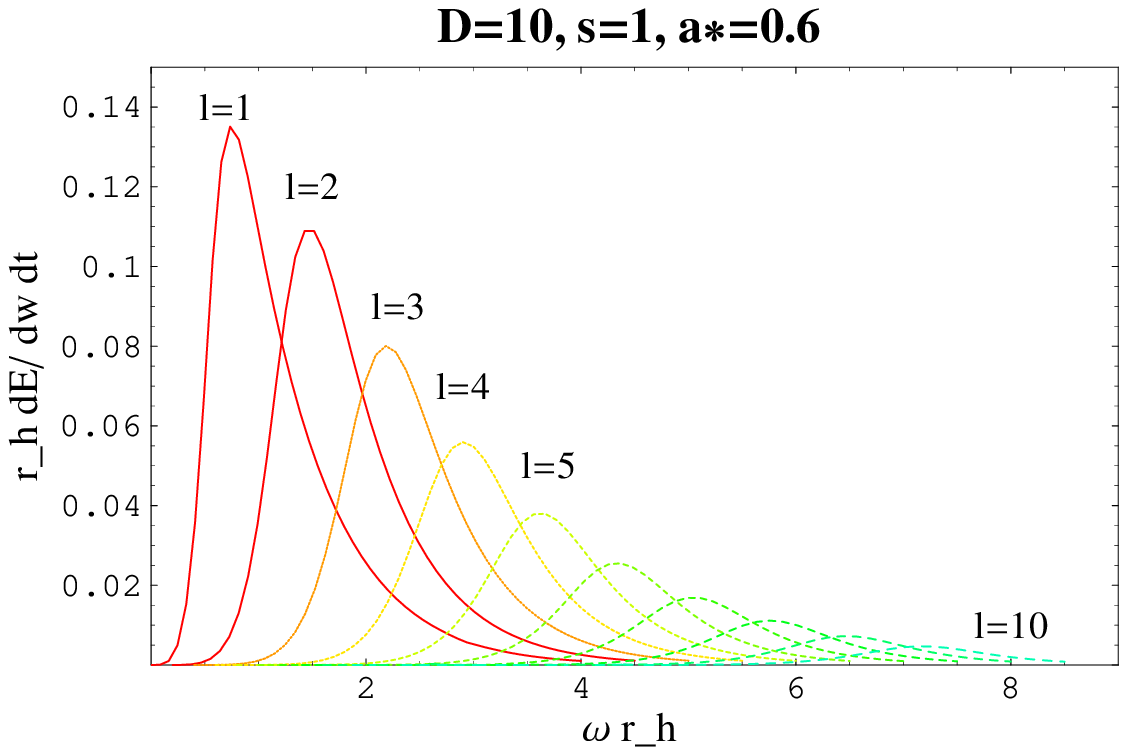}
\includegraphics[width=.43\linewidth]{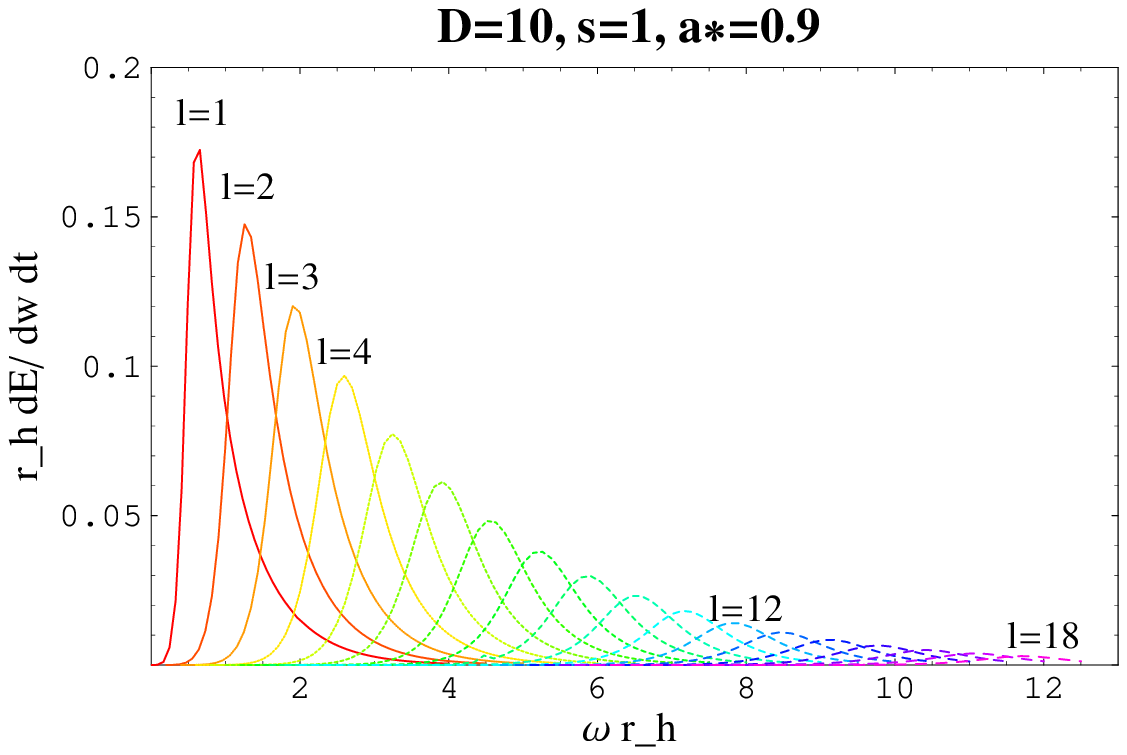}
\includegraphics[width=.43\linewidth]{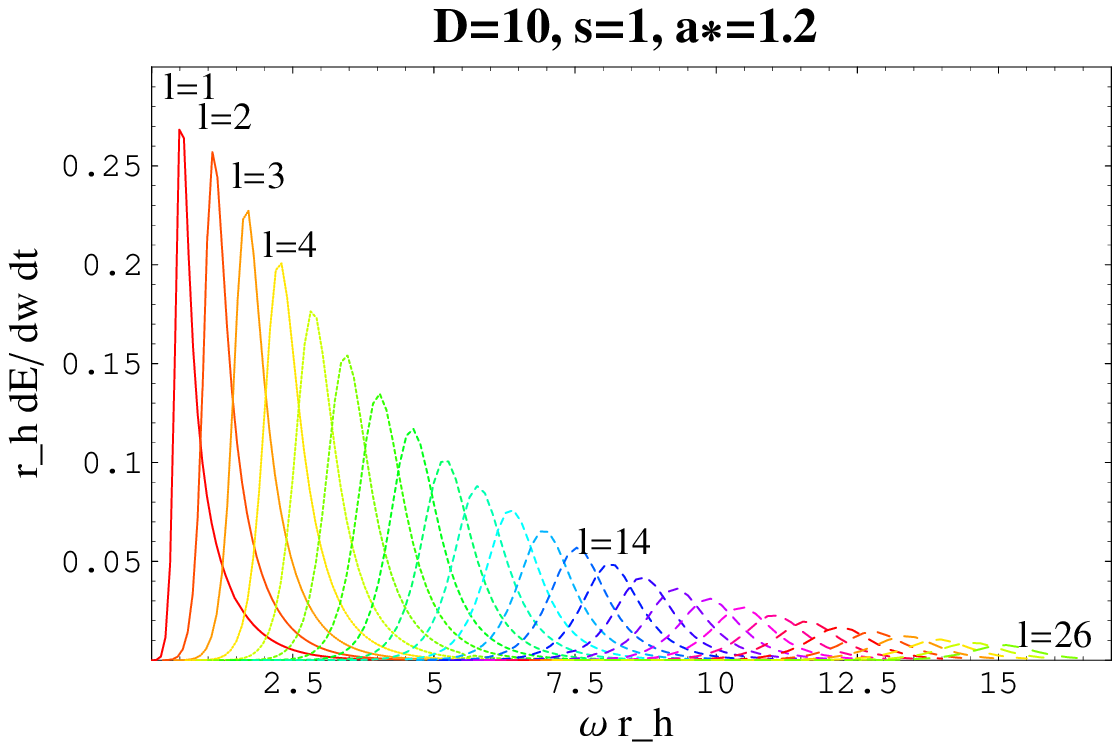}
\includegraphics[width=.43\linewidth]{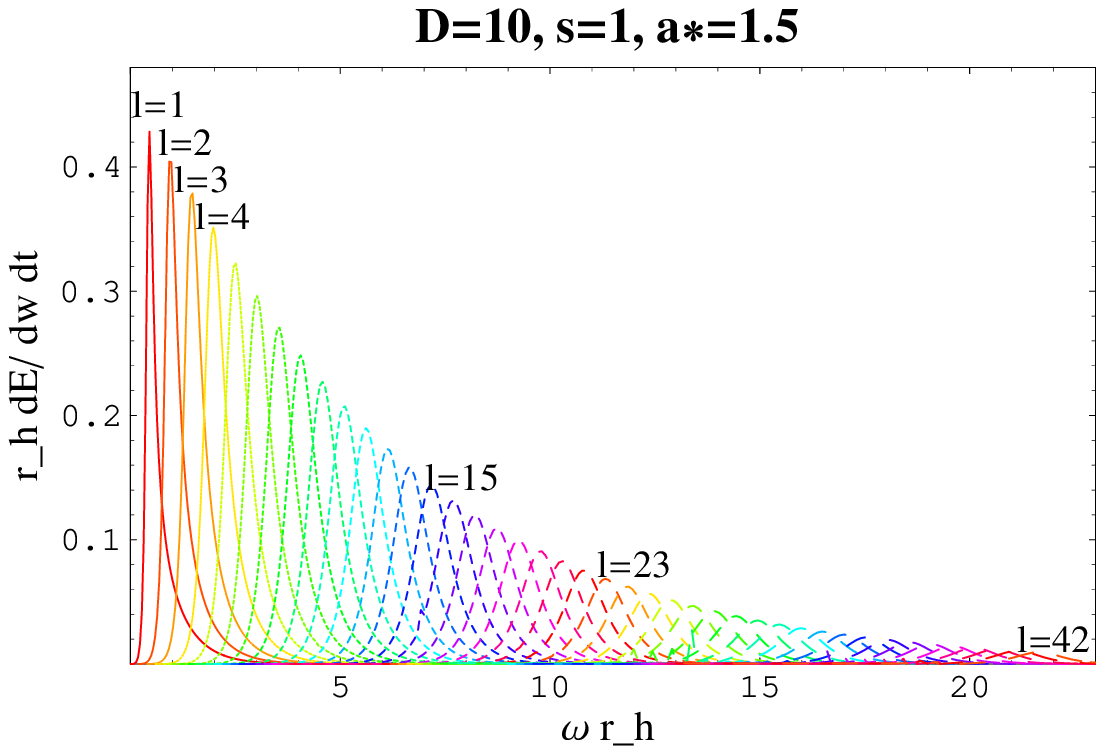}
\end{center}
\caption{Hawking radiation from $D=10$ black hole. $s=1, a=0.01, 0.3, 0.6,0.9, 1.2, 1.5$.}
\end{figure}

\section{Time evolution} From the ratio of energy and angular momentum in eq.\ref{rates}, we
can define a scale invariant function $\gamma(a_s=a/r_s)$ as
follows:
\begin{eqnarray}
\gamma^{-1}(a_s)&\equiv& \frac{d \ln a_s}{d \ln
M}\label{h-function}\\
&=&\frac{n+2}{2}\left(\frac{1}{a}\frac{d J}{ d
M}-\frac{2}{n+1}\right).
\end{eqnarray}

Now we calculate the ratio of final($M_f$) and initial($M_i$) energy
of black hole by integrating the eq.\ref{h-function} with $a_s({\rm
ini})$ for initial angular momentum.
\begin{eqnarray}
\frac{M_f}{M_i}={\rm Exp}\left(\int_{a_s(\rm ini)}^{a_s(\rm final)}
d a_s \frac{\gamma(a_s)}{a_s }\right).
\end{eqnarray}
The amount of energy which is radiated in spin-down phase ($0
\thickapprox a_s({\rm final}) \leqslant a_s \leqslant a_s ({\rm
ini})$) is $(M_i-M_f)$ and $M_f$ will be also radiated in
Schwarzschild phase where the angular momentum of black hole is
zero.

Next, let us consider the evolution of
the black hole. Since time scales as $r_s^{n+3}$ in $(4+n)$
dimensions \footnote{We can easily understand this by simply looking
at the formula $-dM/dt \sim A T^4$ where the surface area of horizon
$A\sim r_s^2$ for brane fields and the temperature of the hole
$T\sim 1/r_s$ and $M\sim r_s^{n+1}$.}, it is convenient to introduce
scale invariant rates for energy and angular momentum as follows.
\begin{eqnarray}
\alpha (a_s) &\equiv& - r_s^{n+3}\frac{d \ln M}{d t},\\
\beta (a_s) &\equiv& - r_s^{n+3}\frac{d\ln J}{d t},
\end{eqnarray}
with these new variables $\gamma(a_s)$ can be written as
$\gamma^{-1}(a_s)= \beta/\alpha(a_s) -(n+2)/(n+1)$. We also
introduce dimensionless variables $y$ and $z$ to take angular
momentum and mass of the hole:
\begin{eqnarray}
y &\equiv& -\ln a_s, \\
z&\equiv& -\ln \frac{M}{M_i},
\end{eqnarray}
then finally we get the time variation of energy and angular
momentum in terms of scale-invariant time parameter
($\tau=r_s^{-n-3}({\rm ini})t$) with initial mass of the hole:
\begin{eqnarray}
\frac{d z}{d y }&=& \frac{\alpha}{\beta-\alpha \left(\frac{n+2}{n+1}\right)}, \nonumber \\
\frac{d y}{d \tau}&=& (\beta- \alpha\left(\frac{n+2}{n+1}\right))
~e^{\frac{n+3}{n+1}z}. \label{DE for z,y}
\end{eqnarray}

After finding the solutions $z(y)$ and $\tau(y)$ of the coupled
differential equations \ref{DE for z,y}, one can get $y(\tau)$ and
$z(\tau)$, hence $a_s$ and $M/M_i$, as a function of time. From
these, one can find how other quantities evolve, such as the area.

Up to now we have used a
unit where the size of event horizon is fixed as $r_h=1$ and angular
momentum of the hole is parameterized by ($a_h=a/r_h$). For
conversion of unit, the following expressions are useful with $a_s =
a_h /(1+a_h^2)^{1/(n+1)}$.
\begin{eqnarray}
\alpha(a_s)&=&- \iota_n^{n+1}(1+a_h^2)^{\frac{2}{n+1}}r_h^2 \frac{d M}{dt},\\
\beta(a_s)&=&-\kappa_n^{n+1}(1+a_h^2)^{\frac{2}{n+1}}r_h^2
\frac{1}{a}\frac{d J}{dt},
\end{eqnarray}
where
\begin{eqnarray}
\iota_n &=& r_s M^{-\frac{1}{n+1}}=\left(\frac{16\pi G}{(n+2)\Omega_{n+2}}\right)^{\frac{1}{n+1}},\\
\kappa_n &=& \iota_n (\frac{n+2}{2})^{\frac{1}{n+1}}.
\end{eqnarray}

\begin{figure}
  \begin{center}
    \includegraphics[width=.42\linewidth]{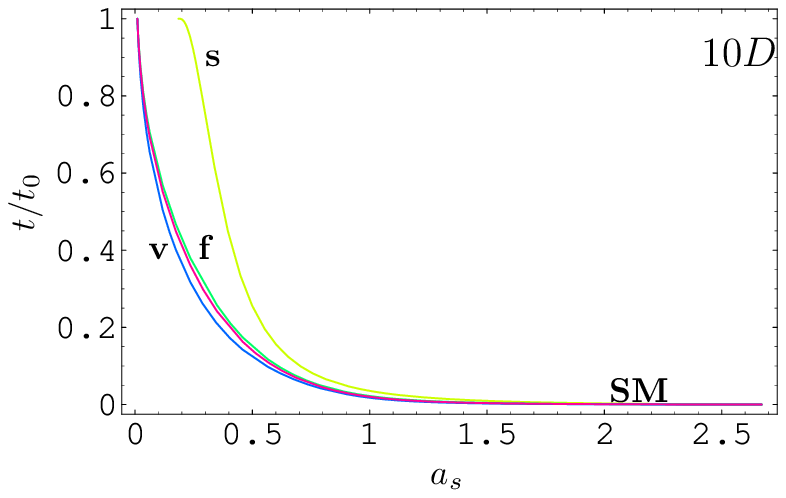}
    \includegraphics[width=.42\linewidth]{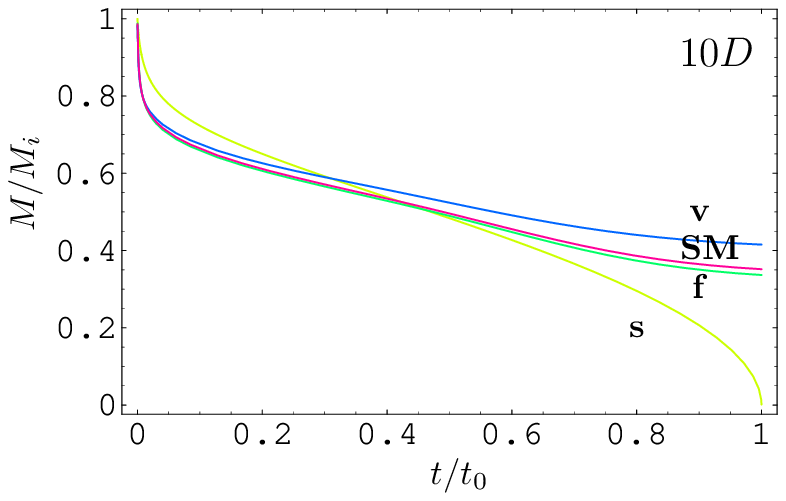}
    \end{center}
  \caption{
    Evolution of bh in $D=10$.
    \label{time_evolution_t_2}}
  \end{figure}

In Fig.\ref{time_evolution_t_2}, black hole evolution in units of the initial mass as a function of rotation parameter
$a_s$ for scalar(s), fermion(f), vector(v), and sum of all the standard model particles(SM) in $D=5$(left) and $D=10$ (right) are shown. The initial angular momentum parameter is fixed by $a_s=0.83$ and $2.67$ in $D=5$ and $D=10$ that are the maximal rotations allowed by the initial collision, respectively. The mass of the hole goes to zero before the rotation parameter goes to zero when only scalar emission is available. However, when all the standard model fields are turned on, the evolution is essentially determined by the spinor and vector radiation. It is found that a black hole spins down quickly at the first stage with large rotation parameter $a_s$ and the decrease of rotation parameter slows down as angular momentum of the hole is reduced.

%
When all the standard model fields are turned on (SM), the evolution is essentially determined by the spinor and vector radiation.
The figures show that a black hole spins down quickly at the first stage with large rotation parameter and the decrease of rotation parameter slows down as angular momentum of the hole is reduced.

\section{Summary and Discussion}
The complete description of Hawking radiation to the brane localized
SM fields and the consequent time
evolution of mini black hole in the context of low energy gravity
scenario has been made.

We have developed analytic and numerical methods to solve the radial
Teukolsky equation which has been generalized to the higher dimension
($D=4+n$). Two main points in our numerical methods are as follows.
First, we have imposed the proper purely-ingoing boundary condition
near the horizon without the growing contamination of the out-going wave
by extracting lower order terms explicitly.
Second, we have developed the method to
fit the in-going and out-going part from the numerically
integrated wave solution at far field region
by explicitly obtaining the next-to-next order expansion
(or next-to-next-to-next order in vector case) of the solution.
With these progress in numerical treatment, we can
safely integrate the generalized Teukolsky equation
up to very large $r$ without out-going wave contamination.

Then we have calculated all the possible modes
to completely determine
the radiation rate of the mass and angular momentum of the hole.
Totally 3407 are computed explicitly, other than the modes which
are confirmed to be negligible.
A black hole tends to lose its angular momentum
at the early stage of evolution.
However the black hole still have a sizable rotating
parameter after radiating half of its mass.
More than $70\%$ or $80\%$ of black hole's mass is lost
during the spin down phase.

Now that we have completely determined the radiation and evolution
of the spin-down and Schwarz\-schild phases, only remaining hurdle
is the evaluation of the balding phase, which is still being disputed
due to its non-purturbative nature,
to extract the quantum gravitational information at the Planck phase from the experimental
data at LHC.

\end{document}